\documentclass{article}
\usepackage{graphicx} 
\usepackage{listings}
\usepackage{url}
\lstnewenvironment{JAVA}{\lstset{language=Java}}{\footnotesize}
\lstnewenvironment{JAVAN}{\lstset{language=Java,numbers=left}}{}

\newcommand{\rep}[1]{#1}

\title{Guided Object-Oriented Development}
\author{Harrie Passier, Lex Bijlsma, Ruurd Kuiper, Kees Huizing}

\begin{document}

\maketitle

\subsection*{Abstract}
To improve the quality of programs we provide an approach to guidance in the process  of program development. 
At the higher level the various activities and their dependencies to structure the process are identified. At the lower level, detailed, practical rules are given for the decision-making in the development steps during these activities. 
The approach concentrates on structure and behavior of a single class. It includes design  and specification and is compatible with methodologies for programming in the large. Informal specifications are introduced to help develop correct and robust code as well as corresponding tests. A strict distinction is made between external design
and specification on one hand and internal design and specification on the other hand, which helps in keeping control over complexity.
The approach also exploits the separation of success and failure scenarios. 
A worked-out example is provided.

\section{Introduction}

\rep{Software often contains errors. This is the case for professionally developed software as well as for the code students produce in programming assignments. This suggests that programming education should pay more attention to correctness issues to prevent mistakes, and to testing to find any remaining ones. In particular, it has been observed that students' tests tend to be far from complete \cite{Edwards2014, Bijlsma-Test}.}

Developing software of high quality is very difficult and is a complex task. For complex tasks, conceptual knowledge is not sufficient: procedural knowledge, guidance in how to apply the concepts effectively, is of great importance \cite{merrienboer2007}.

In the case of software development, examples of conceptual knowledge are the syntax and semantics of programming languages and design notations. Procedural knowledge concerns how to employ these effectively so that the software developed functions correctly, can be maintained and, because it is used in many different contexts and operated by humans, is robust.

Although techniques exist which are known to improve software quality, today’s programmers seem reluctant to use them. 
One reason for this is that techniques for quality assurance and, in particular, software testing
are often isolated topics, placed late in the curriculum for software engineering or general computer science studies~\cite{Christensen2003}. 

In the Erasmus+ programme Quality-focused Programming Educa\-tion {\sc qped}\footnote{\texttt{https://qped-eu.github.io/}} we studied the question of how the teaching of software development can be improved, where we focused on the widely used object-oriented paradigm that provides particularly suitable structure.

In the present paper we introduce the approach we developed: Guided Object-Oriented Development ({\sc good}). 
The approach is fully elaborated in a technical report~\cite{passier2022}.
Noteworthy aspects of the approach are the following.

\begin{itemize}
\item 
\rep{We provide students with a clear road map through the jungle of design steps and decisions, so that they will not be confronted with a blank page without any idea what action to take next. For every step guidance is available explaining how to go about performing it, which decisions have to be taken and what to watch out for}.
\item
\rep{An essential ingredient in the approach is to write explicit and precise specifications of each entity: precise specification is the foundation of the concept of correctness.
The approach leads to well-structured specifications that offer a framework for better readable code and more comprehensive testing.}
\item 
\rep{Robustness is catered for by addressing not just the main, successful, \emph{happy path scenario}, but also the non-happy paths. This again in a well-structured manner, using subspecifications.}
\item
Rather than delegating testing to the penultimate phase of development, the approach enables the integration of writing tests as early as possible in the development process \cite{Bijlsma-Integrated} - and in programming education.
\item
The approach does not prescribe a linear series of steps from requirements to code, but allows specification, design and implementation activities to proceed in many different orders and interleavings. It is therefore flexible in accommodating different styles according to the nature of the problem and the inclination of the developers.
\item
The approach focuses on getting the code in method bodies correct, not on the structure of class hierarchies. It is therefore compatible with, and can be complemented with, design methods such as Unified Process \cite{Kruchten-Rational}. 
\item
There is a strict distinction between the external view of software intended for clients and users, and the internal view providing information for coding \cite{Bijlsma-Model, leino2002data}. The internal specifications constitute a refinement of the external ones \cite{Morris-Laws, Wirth-Refinement}.
\item
Throughout, specifications are expressed in a {\sc jml}-like format \cite{Leavens-Notation}. However, the preconditions and postconditions may be expressed informally in natural language, as in our experience formal specifications create too heavy a burden in introductory programming courses.  For more advanced courses the use of  formal specifications is definitely an option, possibly supported by verification tools such as Dafny \cite{leino2014dafny, leino2023program}.
\end{itemize}

The paper is organized as follows. 
The next section,~\ref{sec:guidance}, describes the {\sc good} approach: the two levels of guidance, and how robustness is incorporated.
Section~\ref{sec:example} contains a worked-out example, development of a Bag, that illustrates the approach.
In section~\ref{sec:tooling} a tool is briefly described that supports {\sc good}. 
The tool specifically enables to organize the various artifacts for the various views as one large artifact from which the tool offers projections as desired.
Related work is discussed in section~\ref{sec:related}.
Lastly, section~\ref{sec:discussion} contains conclusions and some discussion.  
\section{The Guidance}
\label{sec:guidance}

Procedural guidance~\cite{merrienboer2007} provides explicit, concrete, stepwise guidance  to support performing or acquiring skills in some complex activity.

Our Guided Object-Oriented Development ({\sc good}) approach provides specific procedural guidance for development: guidance in program design, specification, coding and testing.
{\sc good} concerns decision points in the development process: what these points are and what to do at these points.
On the one hand, it is at a more general level than a development methodology: there is no fixed order of development steps and there are no strict design or coding requirements.
The guidance thus is flexible in its application and compatible with various methodologies, like Unified Process~\cite{Kruchten-Rational}, and also can support diverse didactic approaches.
On the other hand, it is quite precise and explicit in what the decision points are and provides detailed, rule-like advice, about what to do at these points.
{\sc good}, first, provides guidance at a higher level: it identifies the activities that make up program development, the artifacts these should produce, and the relationships between these.
Secondly, using this structure of activities, {\sc good} provides rules to perform the activities, grouped with the activities.

Section \ref{sec:example} contains a worked-out example, development of a Bag, that illustrates the approach.
To provide some concretization of the ideas already sketched in the present section, we here refer to this example at an intuitive level.

\subsection{The activities structure}

The focus of our approach is the development of a single class with an emphasis on correctness with respect to explicit specifications. It is compatible with methodologies that lead to the design of a software system consisting of cooperating classes.     

The procedural guidance essentially concerns decision points in development: what to do at these points, but also what in fact these points, often left implicit, are.
The guidance also helps navigating between such points, for this we structure the development into activities. 

\begin{figure}[htp]
	\centering
	\includegraphics[scale=0.7]{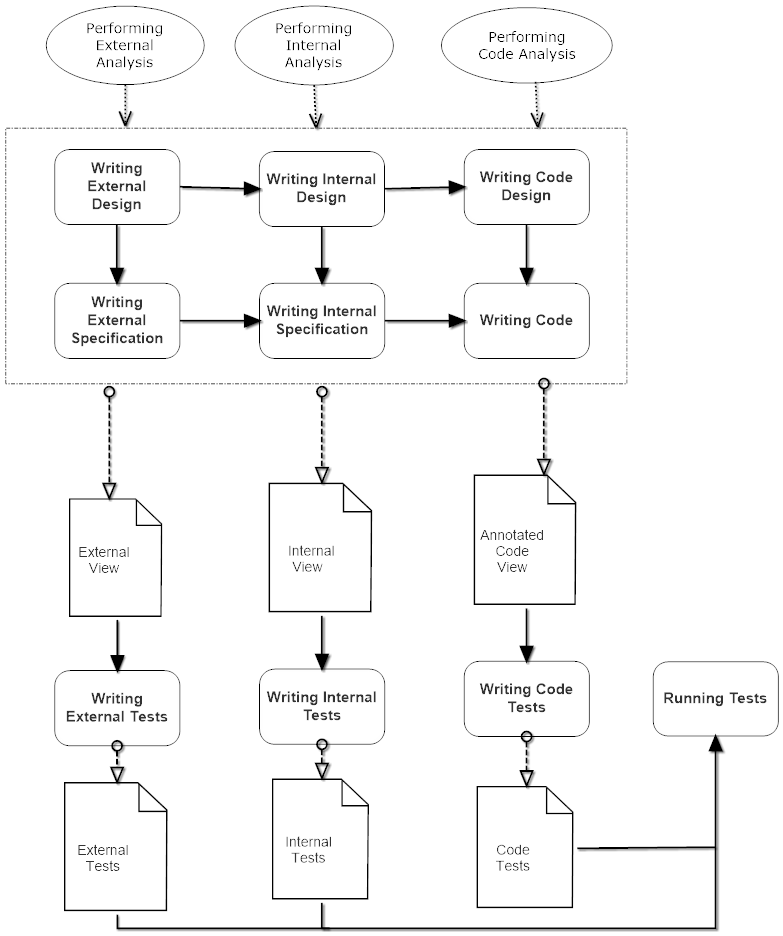}
	\caption{\label{fig:Overview}Overview of the approach}
\end{figure}

Figure \ref{fig:Overview} distinguishes and categorizes the various activities for the development of a program, the artifacts produced in these activities, and the relationships between them that lead from problem assignment to tested implementation.
Every row in the figure corresponds to a particular kind of activity (or the artifact produced), such as writing a design or a specification.
The columns in the figure correspond to stakeholders and their views.
We distinguish three kinds of stakeholders, each of which needs a separate kind of information:

\begin{itemize}
\item 
     External. Information for the users or clients of the software.
\item
     Internal. Information for the high-level developer.
\item 
     Code. Information for the coder.
\end{itemize}

In Figure \ref{fig:Overview}, \emph{activities} are indicated by rounded boxes, e.g.\ Performing External Analysis, Writing Internal Specification. Distinguishing the various activities is important for achieving separation of concerns, and also for accommodating different strategies for ordering development steps.
The activities may result in an \emph{artifact}, e.g. External Design, Internal Specification. In each column, these artifacts are collected in a document, called a \emph{view}, that contains the information relevant to the column's stakeholder, and also forms the basis for writing tests.
Although these views are intended for separate audiences, in practice we find it useful to have all views generated from a single source file, in order to prevent the views from getting out of sync. A dedicated tool \cite{albers2022towards} performs the projection of this common source onto views.

We elaborate a bit on the artifacts. 

There is a strict distinction between the external view, intended for clients and users, and the internal view, intended for developers/programmers \cite{Bijlsma-Model, leino2002data}, thus realizing the principle of \emph{information hiding} \cite{Parnas-Criteria}.
The internal specifications constitute a refinement of the external ones \cite{Morris-Laws, Wirth-Refinement}.

As an example, further elaborated in section~\ref{sec:example}, think of developing a class Bag. 
This is a collection where, different from sets, duplicate elements are represented.
\begin{itemize}
\item
     The External View (also know as the {\sc api}) is the public declarations  and public specifications of the class.
     For the Bag, public methods, like add, remove and display.
\item 
     The Internal View is the extension of the External View with private declarations and private specifications. For the Bag, instance variables, like a -- possibly sorted -- array or linked list to store the elements, and private methods like insert.
\item 
     The Annotated Code View is the extension of the Internal View to a complete, annotated implementation. For the Bag, e.g., implementing the bodies of the public and private methods.
\end{itemize}

All artifacts in these views are developed in three main activities: Performing Analysis (collecting all the information needed to produce the design and specification), Writing Design (in which the structure is determined), and Writing Specification (in which the behavior is described).
The Analysis activities do not directly produce artifacts, but collect insights that are used during the Design and Specification activities.

Additional to the development of a class itself, tests are developed, again corresponding to the views: External (Blackbox) tests, Internal (Greybox) tests, and Code (Whitebox) tests. 
This amounts to translation of the specifications from the views into test cases, which is relatively
straightforward and hence relegated to just one activity for each test: Writing the corresponding Tests.

For the Bag example, tests could be as follows.
\begin{itemize}
\item
     An External Test is that adding and removing elements satisfies the representation of duplicates.
\item 
     An Internal test is that insertion in a sorted array is done appropriately.
\item 
    A Code test could address certain specifics of the sorting algorithm. 
\end{itemize}

In figure 1 the relationships between the various development activities and artifacts are shown.
It is important to realize that this process is not stepwise: the relationships are more flexible and intricate than that.  We distinguish three types of relations.

\begin{figure}[h]
	\centering
	\includegraphics[scale=0.8]{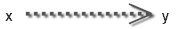}
	\caption{\label{fig:input}x is input for y}
\end{figure}
x \emph{is input for} y indicates that x produces insights and decisions that are input for an activity.
An Analysis activity does not directly produce an artifact: it provides information to the activities that produce the artifacts.
For the Bag example, one needs to analyse what a bag exactly is, and what external interface  is desired.

\begin{figure}[h]
	\centering
	\includegraphics[scale=0.8]{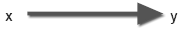}
	\caption{\label{fig:needs}y needs x}
\end{figure}
 y \emph{needs} x indicates that x needs information from y. This means that x and y may evolve together, and going back and forth between (working on) x and y is possible. 
It does also mean that on completing y, all information needed for x needs to be present, i.e., y can only be considered completed when x is completed - which requires checking this in the process.
For the Bag example, for devising tests one need not wait till the corresponding View artifact is  complete, but the tests cannot be considered complete until the View artifact is.

\begin{figure}[h]
	\centering
	\includegraphics[scale=0.8]{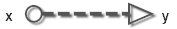}
	\caption{\label{results}x results in y}
\end{figure}
 x \emph{results in} y indicates that an activity results in an artifact. For the Bag example, this means that the specifications, annotated code and tests are available.

The procedural guidance is flexible with respect to the order of steps and their repetition.
This can be exploited in various ways. We give some examples.

 \begin{itemize}
\item 
     Activities can be revisited when insights obtained during development motivate this.
\item   
     Different development methods and didactic approaches, with their own ordering of activities can be accommodated.
\item 
     Programming assignments themselves may suggest an ordering or even a selection of activities.
     There are assignments where the best approach is to first perform the design in all columns before thinking about specifications.
     For example, if the solution heavily leans on the application of a Design Pattern or a specific algorithm.
     There are also assignments where the essential conceptual step is data refinement, in which case it is advisable to fully complete the second column before producing coding.
     The \textsc{good}-approach is amenable to both.
 \end{itemize}

Both levels are further elaborated in the next subsection,~\ref{sec:activities}.

\subsection{The activities and their guidance}
\label{sec:activities}

The second level of guidance identifies, per activity, the decision points that occur and provides guidance as to how to perform steps that produce the artefacts. 
This guidance is given as a set of rules grouped with the activities. The rules are often suggestions or rules of thumb, rather than strict prescriptions.

For the Bag example, think of rules how to choose a sensible external/internal interface, how to decide on instance variables like array or linked list, how to maintain and argue consistency between the artefacts, how to decide on tests, ensure test coverage, execute and interpret tests, and so on.

We distinguish between Design and Specification. Design is about the decomposition of the whole system in modules, a description of the responsibility of each module and the relationship among the modules. As such, Design describes the structural aspects of a system. Because the approach is meant for {\sc oo} systems, the modules are classes with their attributes, methods and associations. Specification on the other hand describes the behavior of the system in terms of pre- and postconditions and invariants on the values of attributes.
What is new is that we distinguish between External Design and Specification as well as Internal Design and Specification, where the internal specification is a refinement of the external specification.
In fact, we combine the results of Liskov and Guttag \cite{LiskovGuttag2000}, which distinguish between design and specification, and the results of Leino and Nelson \cite{leino2002data}, which distinguish between external and internal specifications.
This combination enables to structure guidance to correctly implement the functionality as well as to define external, internal and code tests. 

The specifications we use typically consist of a number of subspecifications, each geared to a separate condition on the input parameters and on the state of the calling environment. This structure helps in developing implementation and tests, as these can deal with each subspecification separately. It also forces the programmer to decide explicitly what level of robustness is appropriate, depending on the extent to which the caller's state is known.

For the Code View, Design concerns the structure of the code: for this, e.g., pseudo code, assertions, information about chosen specific algorithms, where applicable placed in comments, is used.  
The Annotated Code comprises the code (also containing the structure information as in Design) that implements the required behavior and all specifications.

The artifacts can generally be compiled/executed, thus enabling early correctness checks.

Test are constructed fairly directly from the information present in the Views.
Therefore we do not distinguish Design and Specification activities for Tests, but just a Writing Tests activity.

The approach allows the external (blackbox) tests to be developed as soon as the external specification is available. This ensures that these tests do not suffer from misconceptions or unwarranted assumptions introduced later in the process, and can be used as a guide to further development, thereby providing all advantages associated with the test-first ideology. However, we do not insist on all tests being developed upfront, as aspects such as code coverage can only be investigated once the implementation is available \cite{Bijlsma-Integrated}.
We now present a brief description of all the activities. A more detailed description can be found in \cite{passier2022}

\paragraph{External Analysis}
\begin{itemize}
    \item[Goal] Getting information and/or insight in what functionality precisely must be realized and what conditions should be imposed.    
    \item[Input] The starting point is an idea or an assignment about a class with certain functionality. 
    \item[Output] The result is insight needed to be able to design and specify the External View in terms of domain concepts. For the design, information and insight is needed for the class and method names, the names and types of the method's parameters, and a description of the class's and methods' responsibilities. For the specification, information and insight is needed for conditions imposed on the states an instance of the class may be in and on the behavior of methods in terms of pre- and postconditions. This will also give rise to the identification of further classes that represent the domain concepts employed in the description of the current class. 
    \item[Guidance] Study the problem domain, collect and define all relevant domain concepts, describe some examples of use, think of suitable names for the types (!) and methods, think of the minimal data needed for processing, units of values, return values, and constraints on the data.    
\end{itemize}

\paragraph{External Design}
\begin{itemize}
    \item[Goal] Defining the public structure in terms of class and methods signatures, each signature provided with a description of the responsibility of the entity in terms of the domain concepts.  
    \item[Input] Information and/or insights from the External Analysis activity to be able to define a class signature, consisting of a name, and method signatures, each consisting of a method name, parameter names, and suitable parameter and return types.    
    \item[Output] The code structure of a class consisting of the class signature and public method signatures provided with a description of intended meaning of each entity using the tag \texttt{@desc}.
    \item[Guidance] Using the results from the Analysis activity, define clear names of the class (noun), methods (verbs) and parameters (noun) and the types (of parameters and return types).  New types introduced here will have to be realized later as new classes, following the same procedure we are now describing. Characterize the meaning of the class and each method in one sentence in terms of domain concepts gathered during the External Analysis. 
\end{itemize}

\paragraph{External Specification} 
\begin{itemize}
    \item[Goal] Adding conditions on the behavior of the class and public methods.  
    \item[Input] The results of the External Design, i.e. the class and public method signatures including a description of their responsibility, and information and/or insights from the External Analysis activity to be able to define constraints on the behavior of the class and methods. 
    \item[Output] An External View in terms of domain concepts. The class is provided with a description (`class invariant') of the possible object states in terms of domain concepts, using the tag \texttt{@inv}. (Example: each value occurs in a bag with a nonnegative frequency.) The public methods are provided with pre- and postconditions, using the tags \texttt{@requires} and \texttt{@ensures}, and information of their footprint, using the tags \texttt{@pure} and \texttt{@assignable}.   
    \item[Guidance] Are there limitations on the states an instance may have leading to one or more class invariants? For each method, are there requirements on each parameter or combinations of parameters? In which state can a method be called? What return values are possible? How does the method affect the state of the object? What does the method change within this or other objects?      
\end{itemize}

\paragraph{External Test}
We define External Tests on the basis of the External View. 
External tests can thus be written prior to implementing the functional code. The benefit is that the External View is translated stepwise into an External Test instead of a test functioning as (partial) specification, as in {\sc tdd} \cite{beck2003test}. 
This results in a test with a high test coverage. The test code can only be executed after the present class has been implemented as well as the other classes mentioned in the signatures.  The latter can be replaced by stubs if needed.  
\begin{itemize}
    \item[Goal] To write tests of the class' functionality based on the External View of the class which will allows to check the correctness of any chosen implementation.  
    \item[Input] An External View consisting of a design and specifications.  
    \item[Output] A test with for each combination `method \texttt{M} - (sub)specification \texttt{C}' a test method with name \texttt{testMC} containing the assertions needed according to the specification provided with annotation \texttt{@Test}.
     \item[Guidance] The specifications are translated into tests following three steps: 1) make a test model, 2) choose coverage criteria, and 3) design test cases. Several test models exist, for example Equivalent Classes and Boundaries of input variables, Decision Tables and Mutation testing \cite{jorgensen2013software}. For practical reasons, we have chosen for Equivalent Classes and Boundaries of input domain variables and are able to find sufficient test cases in most situations corresponding to the specification. In the next step a coverage criterion is chosen and applied. Again for practical reasons, we choose for the All combinations coverage. The second step results in all the (combination of) input values needed. In the third step, the test methods containing the assertions are established combining the input values with the expected output values.
\end{itemize}

\paragraph{Internal Analysis}
In the Internal Analysis we try to gain insight into how to represent the external domain concepts into a suitable internal representation in terms of attributes and private methods. What is suitable is determined by algorithm efficiency, practices of coding, choices previously made, and available Java types. Although this can be a trivial activity, for example when a method has to manipulate integer values a type \texttt{int} is obvious, this can be a complex activity with many options to choose from.  

\begin{itemize}
    \item[Goal] To get information and ideas needed to be able to choose and/or define a suitable internal representation of the external domain concepts in terms of attributes and private methods.
    \item[Input] This analysis activity requires knowledge of the standard Java types and general class design and specification in cases a new type is needed. Insights gained from the External Analysis can help with this. Furthermore, choices made in the External Design and Specification can impose constraints on the representation that can be chosen.
    \item[Output] The result is information and/or insight needed to be able to design and specify the present class in terms of program entities, i.e.\ the class attributes and methods. 
    \item[Guidance]  To emphasize an External View is independent of the chosen internal representation, thinking of the External View as an interface instead of a class can help, i.e.\ several implementations may be possible. Private (helper) methods avoid code duplication in cases where a same computation will take place in several  locations. Consider whether a value should be stored in an attribute or whether it can be calculated just-in-time when needed. If one foresees a method or a class with two or more responsibilities, i.e.\ low cohesion, it is better to split the method or class into separate entities. This will result into helper objects and - methods. Sometimes the functionality of the helper-object can be found in an existing Java library, but in case a helper object needs to be constructed from scratch, a new class has to be introduced and for which one has to follow the procedure again.        
\end{itemize}

\paragraph{Internal Design}
\begin{itemize}
    \item[Goal] To decide on the types and names of the class' attributes and to decide on the responsibility, signature, and return type of each private method.   The types here are types available in the programming language, in its libraries, or new types to be constructed (or already constructed) in the present project.
    \item[Input]  Information and insight from the Internal Analysis activity and the External Design.  
    \item[Output] An internal class design with attributes (names and types) and private methods (signatures and return types)  supplemented with a description of responsibility using the \texttt{@desc} tag.
    \item[Guidance] Here, it is determined what private attributes with getters and private methods are certainly needed. We have to be careful with adding attribute setters. In choosing an internal representation, it is important to realize that an External Design makes more internal representations possible. Each class entity is provided with a description, using the \texttt{@desc} tag, to explain their intended meaning which will be elaborated on during the specification activities to make them complete and precise. 
\end{itemize}

\paragraph{Internal Specification}
\begin{itemize}
    \item[Goal] Determine how the domain concepts, described in the External View, correspond to the class attributes described in the Internal Design. Furthermore, conditions are added on the behavior of the class' attributes and private methods. External specifications already present will be reformulated (in fact, refined) into programming language entities. The Internal specification is meant for the implementer of the class, clients should not be dependent on it  in the sense that changes in the internal specification only will not necessitate changes in client code (information hiding).
    
    \item[Input] First, the output of the Internal Design, i.e. the class' attributes and private methods.
     Secondly, information and/or insights from the Internal Analysis activity to be able to define constraints on the behavior. Lastly, the output of the External Specification.   
    
    \item[Output] An Internal View in terms of class attributes and private methods. The class is provided with a description of the states the class may be in using the tag \texttt{@inv} describing an invariant.
    The methods are provided with pre- and post conditions, using the tags \texttt{@requires} and \texttt{@ensures}, and information of their footprint, using the tags \texttt{@pure} and \texttt{assignable}. Extra, compared to the External View, is the tag \texttt{@represents} describing how the domain concepts, used in the External View, correspond to the class attributes. This clause is called the \emph{representation relation} and plays an important role in proof rules. 
 
    \item[Guidance] The combinations of attribute values that are allowed and which not should be determined. The class invariant \texttt{@inv} is the place to record these limitations and must be respected by every method of the class, i.e. each invariant is silently added to all the methods' preconditions and postconditions.  
    
    In the end, The Internal Specifications should be a refinement of the External Specifications which means that it should satisfy three rules: (1) the representation relation and the internal class invariant should together imply the external class invariant, (2) for every method, the representation relation and external precondition should together imply the internal precondition, (3) for every method, the representation relation and internal postcondition should together imply the external postcondition.

    Sometimes, extra class invariants have to be added to limit the range of values of an attribute or combinations of attributes. For instance, representing the domain concept `month' by an integer $m$ gives rise to the class invariant $1\le m \le 12$.

    In situations where several cases have to be distinguished, a specification can be divided into several subspecifications. We use subspecifications explicitly to differentiate between happy path and non happy path (robustness) behaviour (see Section \ref{overview.robustness}). These can be useful too when the description of happy path behaviour can be separated into multiple subspecifications. As an example, in a phone application we can differentiate between national and foreign telephone numbers where both have own requirements on the phone number. Subspecification mostly occurs in the Internal Specification activity, because here the need for case analysis often appears, but can be origin as part of the External View activities too.           
 \end{itemize}

\paragraph{Internal Tests}
The Internal View replaces with more detail the External View, which can result in extra testcases.    

\begin{itemize}
    \item[Goal] To be able to check the code's functionality against the Internal Specification of the class independent of the chosen Coding.  
    \item[Input] An Internal View consisting of a design and specifications.  
    \item[Output] Extra test cases using the same layout and tags as used in the External Test. 
     \item[Guidance] Extra test cases can origin due to the representation relation as this relation can, for example, limit the values of one or more attributes and/or strengthen or refine the specifications of methods. 

     Also, extra test cases can be needed due to class invariants limiting the range of values attributes can have, leading to new boundary values.  

     Extra test cases can be needed when pre- and postconditions are changed due to the representation relation and/or class invariants. Furthermore, a developer can choose to weaken the precondition and/or the strengthen the postcondition.   

     To test whether a class invariant holds after a method has been executed, private attributes can be reached using extra getters. As an alternative, a dedicated method can be added that checks whether the invariant directly.  
     
     Private methods can be tested declaring them temporary with access modifier package or using extra boolean methods testing the corresponding postcondition after finishing the private method. This boolean method can be called directly after calling the client method that calls the private method. Lastly, one can consider the functioning of a private method as part of the calling client method which is tested.
\end{itemize}

\paragraph{Annotated Code View activities}
We keep the treatment of this topic terse, because it has been covered in detail in programming courses. During the coding activities, based on the Internal View describing the structure the required behavior of a class, the methods are provided with a body. 

During the \textit{Code Analysis activity}, it makes sense to check whether each method so far has one responsibility and to split it up in case of low cohesion. Furthermore, one can think of private helper methods to prevent for, for example, redundant code fragments. Generally, ideas are collected of how the specified functionality can be achieved. Sometimes, existing algorithms or even \textsc{api}'s can be used. In case of a new algorithm, pen and paper elaborations, how to come from input to output, can provide great insight.

In \textit{Code Design}, a method's body structure is developed.
Here, the subspecifications often can help in structuring the method's body, for example to determine firstly if there is a case of exceptional behaviour before a happy path scenario can be performed.

In case of a known or a new algorithm, meaningful local variables and, if needed, extra (private) helper methods are chosen. 

In case of a complex algorithm, a first design-step can be to describe the algorithm in natural language or pseudo code. These text lines can function as inline comments later. In some cases they can be retained as assertions that will contribute to the Code Tests. 

Sometimes, the chosen algorithm can't be efficiently implemented with the chosen attributes during the previous activity. In such a case an alternative data representation should be considered.       

During \textit{Writing Code}, the Code Design is translated into executable programming code resulting in what is called the Annotated Code View, consisting of the Code, the specifications and inline comments. In the end, the final Annotated Code View contains both the External View and the Internal view.      

During coding, one have to search for duplication in code, which can be solved by extra helper methods, which is a design decision. The code must satisfy the design and specifications which can be checked manually during this activity Coding. Based on the views, tests are developed to check on correctness automatically later. One should avoid all kinds of common made bugs in code as null values and array-access out of bounds.      

\paragraph{Code Test}
Code testing, also structural or whitebox testing, uses the code's structure as information to define test cases. The standard approach is to apply some coverage criteria~\cite{Ghezzi2002} as Statement coverage and Edge coverage and to search for special code constructions that can fail. Finally, one has to check on check for duplicate test cases already part of the External or Internal Tests.   

In our approach, we consider code testing as foreknowledge and refer to the standard literature~\cite{BinderOOTesting}.

\subsection{Robustness}
\label{overview.robustness}
So far, we assumed that methods are called with arguments satisfying the precondition of a (sub)specification. In situations where the caller's behavior can't be controlled, it is possible that the precondition is not satisfied and as a consequence the postcondition may not be reached. In such situations, a method can be made robust, meaning that regardless the input values a postcondition is always reached.   

To achieve robustness, one has to add subspecifications describing the wrong (combination of) input values and how the program behaves on these wrong input. These subspecifications describe in the \texttt{@requires} tag a situation of erroneous input and describe using the tag \texttt{@signals} the type of exception thrown and the message it will contain. Other choices are sometimes possible, instead of throwing an exception one can return a special value.     

Making software robust, we have to perform potentially all the activities again. For example, as part of the analysis we have to consider the best form of robustness, i.e. throwing an exception, returning a special value, or handle the situation locally. As part of the design, we have to extend the signature of the method with additional information and have to choose a concrete exception type. As part of the implementation, we have to add a suitable control flow to detect erroneous input and to throw the right exception, before the happy-path situation can be processed. The subspecificatons can be valuable as inline comments to structure the implementation. 

Finally, each subspecification describing robustness behavior should be translated into testcases applying the boundary value and equivalent class techniques again. 

We note that it is possible to aim for robustness from the beginning and take undesired inputs into account in all activities in one pass. However, for the sake of separating complexities, it is often better, especially for novice programmers, to complete the happy-path scenarios first and then add robustness as a separate aspect in a second pass. In the artifacts, robustness remains clearly separated from the happy-path specifications, implementation and test cases.

\section{Example}\label{sec:example}
We demonstrate the various activities and artifacts by the example of a class Bag.       

\subsubsection*{External analysis}
A bag differs from a set because it can contain duplicate elements; it differs from a list because it does not have order.
Mathematically, a bag is identified with a mapping from its element type to the natural numbers. For a bag $B$ and element $a$, we therefore write $B(a)$ for the number of occurrences of $a$ in $B$. Throughout, we limit ourselves to finite bags: the methods are designed to preserve this condition. The number of elements in the bag will we call its size.

\subsubsection*{External design} 
\begin{lstlisting}
/**
 * This class represents a bag of elements.
 */
public class Bag<T> {

  /**
   * @desc A new bag is instantiated, with size = 0.  
   */
   public Bag<T>()

  /**
   * @desc Adds an element.  
   */
   public void add(T elem)

  /**
   * @desc Removes one instance of elem 
   */
   public void remove(T elem)

  /**
   * @desc Removes all instances of elem 
   */
   public void removeAll(T elem)

  /**
   * @desc Counts the number of occurences of elem.  
   * @pure
   */
   public int mult(T elem)

  /**
   * @desc Counts the number of elems in the bag. 
   * @pure
   */
   public int size()

\end{lstlisting}

\subsubsection*{External specification} 
\begin{lstlisting}
/**
 * This class represents a bag B of elements of type T.
 */
public class Bag<T> {

  /**
   * @desc A new empty bag is created.    
   * @ensures \forall e: T :: B(e) = 0
   */
   public Bag<T>()

  /**
   * @desc Adds an element.
   * @ensures B(elem) = \old(B(elem)) + 1
   * @ensures \forall e: T | e != elem :: B(e) = \old(B(e))  
   */
   public void add(T elem)

  /**
   * @desc Removes one instance of elem 
   * @requires B(elem) > 0
   * @ensures B(elem) = \old(B(elem)) - 1
   * @ensures \forall e: T | e != elem :: B(e) = \old(B(e))  
   */
   public void remove(T elem)

  /**
   * @desc Removes all instances of elem 
   * @ensures B(elem) = 0
   * @ensures \forall e: T | e != elem :: B(e) = \old(B(e))
   */
   public void removeAll(T elem)

  /**
   * @desc Counts the number of occurences of elem.
   * @pure
   * @ensures mult(elem) = B(elem)
   */
   public int mult(T elem)

  /**
   * @desc Counts the number of elems in the bag.
   * @pure
   * @ensures size() = \sum e: T :: B(e)
   */
   public int size()
\end{lstlisting}

\subsubsection*{Internal analysis}
We have no information about the component type {\tt T}, therefore we cannot employ any kind of search tree to increase efficiency. The simplest choice to represent a bag is using a list.

\subsubsection*{Internal design} 
\begin{lstlisting}
/**
 * @desc This class represents a bag of elements of type T.
 */
public class Bag<T> {

  /**
   * @desc lst contains all elements of the bag with their proper 
   * frequencies in an irrelevant order. 
   */
   private List<T> lst;


  /**
   * @desc A new bag is instantiated, with size = 0.  
   * @assignable lst
   */
   public Bag<T>() {

   }

  /**
   * @desc Adds an element.  
   */
   public void add(T elem)

  /**
   * @desc Removes one instance of elem 
   */
   public void remove(T elem)

  /**
   * @desc Removes all instances of elem 
   */
   public void removeAll(T elem)

  /**
   * @desc Counts the number of occurences of elem.  
   * @pure
   */
   public int freq(T elem)

  /**
   * @desc Counts the number of elems in the bag.  
   * @pure
   */
   public int size()
\end{lstlisting}

\subsubsection*{Internal specification}

That the proof obligations are fulfilled is quite straightforward, because the relation between the bag and the list is quite simple. For example, for the constructor, the size of the list is clearly the size of the bag and for the add operation, adding an element to the end of the list increases its frequency by one and leaves the frequency of the other elements as is.
\begin{lstlisting}
/**
 * This class represents a bag of elements of type T.
 */
public class Bag<T> {

  /**
   * @desc lst contains all elements of the bag with their proper 
   * multiplicities in an irrelevant order. 
   * @represents \forall e : T :: B(e) = (\num_of i: int | 
   *     0 <= i < lst.size() :: lst[i] = e)
   */
   private List<T> lst;

  /**
   * @desc A new bag is instantiated, with size = 0.  
   * @ensures size() = 0.  
   */
   public Bag<T>()
   /**
    * @ensures lst.size() = 0 
    */

  /**
   * @desc Adds an element.
   * @ensures B(elem) = \old(B(elem)) + 1
   * @ensures \forall e: T | e != elem :: B(e) = \old(B(e))  
   */
   public void add(T elem)
   /**
    * @desc Adds an element to the list. In order to keep the 
    * specification simple and readable, we add the element 
    * to the end of the list, although the external 
    * specification and the representation rule do 
    * not force this. 
    * @assignable lst
    * @ensures \forall i | 0 <= i < \old(lst.size()) :: 
    *     list[i] = \old(lst[i])
    * @ensures lst.size() = \old(lst.size()) + 1
    * @ensures lst[\old(lst.size())] = elem
    */

  /**
   * @desc Removes one instance of elem 
   * @requires B(elem) > 0
   * @ensures B(elem) = \old(B(elem)) - 1
   * @ensures \forall e: T | e != elem :: B(e) = \old(B(e))  
   */
   public void remove(T elem)
   /**
    * @desc Removes one occurrence of elem 
    * @requires mult(elem) > 0
    * @assignable lst
    * @ensures mult(elem) = \old(mult(elem) - 1
    * @ensures \forall e: T | e != elem :: mult(e) = \old(mult(e))
    * @ensures lst.size() = \old(lst.size()) - 1 
    */

  /**
   * @desc Removes all instances of elem
   * @ensures B(elem) = 0
   * @ensures \forall e: T | e != elem :: B(e) = \old(B(e))
   */
   public void removeAll(T elem)
   /**
    * @assignable lst
    * @ensures ! mult(elem) = 0
    * @ensures \forall e: T | e != elem :: mult(e) = \old(mult(e))
    * @ensures lst.size() =  \old(lst.size) - \old(mult(elem)) 
    */

  /**
   * @desc Counts the number of occurences of elem.
   * @pure
   * @ensures mult(elem) = B(elem)
   */
   public int mult(T elem)
   /**
    * @ensures \result = \num_of i: int | 0 <= i < lst.size() :: 
    *   lst[i] = elem
    */

  /**
   * @desc Counts the total number of occurrences of all elements 
   * in the bag.
   * @pure
   * @ensures size() = \sum e: T :: B(e)
   */
   public int size()
   /**
    * @ensures size() = lst.size()
    */


\end{lstlisting}

\subsubsection*{Code analysis} 
We choose to implement the List with an ArrayList. 
The complete code with specifications and implementations is as follows.

\begin{lstlisting}
import java.util.ArrayList;
import java.util.List;

/**
 * This class represents a bag of elements of type T.
 */
public class Bag<T> {

  /**
   * @desc lst contains all elements of the bag with their proper 
   * multiplicities in an irrelevant order. 
   * @represents \forall e : T :: B(e) = (\num_of i: int | 
   *    0 <= i < lst.size() :: lst[i] = e)
   */

  
  /**
   * @desc A new bag is instantiated, with size = 0.  
   * @ensures size() = 0.  
   */
   public Bag<T>()
   /**
    * @ensures lst.size() = 0 
    */
   {
     lst = new ArrayList<T>();
   }

  /**
   * @desc Adds an element.
   * @ensures B(elem) = \old(B(elem)) + 1
   * @ensures \forall e: T | e != elem :: B(e) = \old(B(e))  
   */
   public void add(T elem)
   /**
    * @desc Adds an element to the list. 
    * In order to keep the specification simple and readable, 
    * we add the element to the end of the list, 
    * although the external specification 
    * and the representation rule do not force this. 
    * Hence the internal specification is a proper refinement.
    * @assigns lst
    * @ensures \forall i | 0 <= i < \old(lst.size()) :: 
    *    list[i] = \old(lst[i])
    * @ensures lst.size() = \old(lst.size()) + 1
    * @ensures lst[\old(lst.size())] = elem
    */
   {
     lst.add(elem); 
   }

  /**
   * @desc Removes one instance of elem 
   * @requires B(elem) > 0
   * @ensures B(elem) = \old(B(elem)) - 1
   * @ensures \forall e: T | e != elem :: B(e) = \old(B(e))  
   */
   public void remove(T elem)
   /**
    * @desc Removes one occurrence of elem 
    * @requires mult(elem) > 0
    * @assigns lst
    * @ensures mult(elem) = \old(mult(elem) - 1
    * @ensures \forall e: T | e != elem :: mult(e) = \old(mult(e))
    * @ensures lst.size() = \old(lst.size()) - 1 
    */
   {
     int here = lst.indexOf(elem);
     lst.remove(here);
   }

  /**
   * @desc Removes all instances of elem
   * @ensures B(elem) = 0
   * @ensures \forall e: T | e != elem :: B(e) = \old(B(e))
   */
   public void removeAll(T elem)
   /**
    * @assigns lst
    * @ensures ! mult(elem) = 0
    * @ensures \forall e: T | e != elem :: mult(e) = \old(mult(e))
    * @ensures lst.size() =  \old(lst.size) - \old(mult(elem)) 
    */
   {
     while (lst.contains(elem)) {
       lst.remove(elem);
     }
   }

 /**
   * @desc Counts the number of occurences of elem.
   * @pure
   * @ensures mult(elem) = B(elem)
   */
   public int mult(T elem)
   /**
    * @ensures \result = \num_of i: int | 0 <= i < lst.size() :: 
    *   lst[i] = elem
    */
    {
      int count = 0;
      for (T e : lst) {
        if (e.equals(elem)) {
          count++;
        }
      }
      return count;
    }

  /**
   * @desc Counts the total number of occurrences 
   * of all elems in the bag.
   * @pure
   * @ensures size() = \sum e: T :: B(e)
   */
   public int size()
   /**
    * @ensures size() = lst.size()
    */
  {
    return lst.size();
  }
}
\end{lstlisting}

\subsubsection*{Robustness}
We shall make one method, \lstinline{remove}, robust by catering for the possibility that it is accidentally called on an element that does not occur.
\begin{lstlisting}
  /**
   * @desc Removes one instance of elem 
   * @sub elem is present { 
   *   @requires B(elem) > 0
   *   @ensures B(elem) = \old(B(elem)) - 1
   *   @ensures \forall e: T | e != elem :: B(e) = \old(B(e))  
   * }
   * @sub elem is not present { 
   *   @requires B(elem) = 0
   *   @ensures \forall e: T :: B(e) = \old(B(e))
   *   @signals ArgumentNotFoundException("Elem is not present")
   *   }
   */
   public void remove(T elem)
   /**
    * @desc Removes one occurrence of elem 
    * @sub elem is present { 
    *   @requires mult(elem) > 0
    *   @ensures mult(elem) = \old(mult(elem)) - 1
    *   @ensures \forall e: T| e != elem :: mult(e) = \old(mult(e))  
    * }
    * @sub elem is not present { 
    *   @requires mult(elem) = 0
    *   @ensures \forall e: T :: mult(e) = \old(mult(e))
    *   @signals ArgumentNotFoundException("Elem is not present")
    *   }
    */
\end{lstlisting}

\subsubsection*{Tests} 
Due to limited space, we only show how the external test cases can be derived from the External Design and External Specifications concerning method {\tt remove}. For practical reasons, we apply the models Boundary values and Equivalent classes of the input domain variables. Assuming a bag $b1 = \{2, 2, 6, 4\}$, with all combination coverage we get the following test cases.\\

\noindent
For the \textit{happy path}, we choose \lstinline{b1.remove(6)} as a boundary, then we expect \lstinline{b1.mult(6) ==0} and \lstinline{b1.mult(2) ==2} and \lstinline{b1.mult(4) ==1}. 
Assuming $b1 = \{2, 2, 6, 4\}$ again, we choose \lstinline{b1.remove(2)} as a value of the equivalent class, then we expect \lstinline{b1.mult(2) ==1} and \lstinline{b1.mult(4) ==1} and \lstinline{b1.mult(6) ==1}. \\

\noindent
For the \textit{non-happy path}, we choose \lstinline{b1.remove(10)} resulting in the \\ \lstinline{ArgumentNotFoundException} exception as expected value.

\subsection*{Remark}
As argued before, the order of the development activities is flexible. In the development order followed above, columns are completed up to test development one after the other before starting on writing tests. However, the external tests only rely on the first column so these could have been developed at an earlier stage. Similarly, the robustness for the one method could have been addressed earlier as well.

\section{Tooling}\label{sec:tooling}
The views we have introduced use a {\sc jml}-influenced annotation language to attach specific information to each part of the code. The
management and correct usage of these annotations has shown
to be cumbersome, in particular in ensuring that the various views remain in sync when changes are introduced. To overcome this problem, a tool was developed \cite{albers2022towards} that produces the various views as projections from a master document. The tool aims to guide developers and students through highlighted annotations, view-specific information,
warnings and errors, and the generation of test cases. To maximize
its usability, the tool is distributed in the form of an add-on to the
popular development environment Visual Studio Code.
\section{Related work}
\label{sec:related}

Parnas has described in detail the role of software documentation, how it can be produced, and how it can be used \cite{parnas2011precise}. Specifications are important parts of software documentation. Among the benefits are correct implementation and more effective testing. 

In the past, many suggestions have been made to improve the quality of students' software by emphasizing specifications.
One such proposal is the \emph{Refinement calculus} proposed by Morgan \cite{morgan1990programming}, Morris \cite{Morris-Laws} and others.
Here an initial specification undergoes stepwise decomposition into subtasks that are themselves either specified or implemented; the process ends when no unimplemented subtasks remain.
A disadvantage of this approach is that a great deal of pen-and-paper argumentation is necessary to produce even the simplest programs, which discourages its use as a feasible strategy in programming courses.

Extensive experience, on the other hand, exists with Dijkstra's \emph{Program derivation} method \cite{dijkstra1976discipline,dijkstra1988method,gries1981science}.
Here a given specification, consisting of a pre- and postcondition pair in first-order logic, is used as the starting point for a calculation with predicates, based on proof rules for the programming language constructs.
The main creative step in such a calculation is the choice of a loop invariant, for which several heuristics are available.
For at least a decade, first- and second-year students at Eindhoven University of Technology exclusively worked in this format\footnote{During this period author Bijlsma was one of the teachers involved.} \cite{perrenet2005exploring}.
This resulted in programs in the \emph{guarded command} notation, designed on purpose to discourage running or testing the programs: the language contained no input or output statements and no procedure calls and the programs were inherently nondeterministic.
The underlying thought was that programs should possess guaranteed correctness through the method of their construction: programs and their correctness proofs were to be developed hand in hand, with the proof leading the way.
The experience with teaching this method showed that it was indeed possible to train students in applying this method, but that in general they found it too burdensome to follow its prescriptions in real-life programming tasks.
The Dafny system \cite{leino2014dafny} aims to take over some of the proof burden in this system through employment of an automated theorem prover.
Gegg-Harrison et al.\ \cite{gegg2003studying} employ the same semantics; however, they do not apply these to program derivation but to ex-posteriori correctness proofs.
Platzer \cite{platzer2013teaching} uses a similar approach in the context of cyber-physical systems, using the argument that people bet their lives on such systems, so they had better be correct.

Apart from their use in constructing the program itself, specifications are also useful for providing test cases.
Cheon and Leavens \cite{cheon2002simple} propose a testing system based on cooperation of {\sc jml} \cite{Leavens-Notation} and JUnit \cite{gulati2017java}, where the tester need only provide the test case input values.
No prediction of the expected result is required: the {\sc jml} specification simply functions as a \emph{test oracle} by checking the state actually produced against the specification.
This only works if the specification is both formal and complete. 
However, as {\sc jml} does not offer a full range of mathematical concepts, the expressiveness of purely formal {\sc jml} specifications is limited; hence formality and completeness are not easily reconcilable.
Examples of this approach in action are given by du Bousquet et al.\ \cite{bousquet2004case} and Liu \cite{liu2007integrating}.

\emph{Design by Contract} is the specification style advocated by Meyer \cite{meyer2002design}.
In contrast to {sc jml}, this style does not focus on unit tests and does not require test cases to be specified in advance.
Instead, programs are adorned with inline assertions and there is no separate test class.
In the original view, these assertions are merely boolean expressions (in the Eiffel programming language \cite{meyer1997construction}), so without quantifiers or nontrivial mathematics, and necessarily formal.
These assertions are left in the code and are automatically tested at each run of the program.
Paige and Ostroff \cite{paige2004specification} consider its use as a way to get acquainted with formal methods.
Polikarpova et al.\ \cite{polikarpova2013good} propose to extend the expression language used in Design by Contract with references to simple mathematical structures such as sequences.
Experiments show that testing against such `strong specifications' detects twice as many bugs as standard contracts, with a reasonable overhead in terms of annotation burden and run-time performance while testing.
One other disadvantage of the restriction to executable assertions remains: it excludes information hiding, so Liskov's substitution principle \cite{liskov1994behavioral} does not apply \cite{leino2002data}.
Indeed, in Eiffel inheritance does not imply subtyping.

Liskov and Guttag develop classes and methods in a way resembling our proposal~\cite{LiskovGuttag2000,liskov1977}.
The starting point for their treatment of data refinement goes back to the work of Tony Hoare~\cite{hoare1972}. In {\sc good} the same type of data refinement governs the relation between external and internal views, but this terminology is not used by Liskov and Guttag. 
An informal notation is used to describe precondition/postconditions pairs. Differences are that exceptions are not described in a separate tag, but are included in the postcondition tag \textsc{effects} and where we use a \textsc{jml}-based notation, a custom notation has been chosen by them. 
A more fundamental difference is that they limit the \emph{relation} between model entities and implementation variables to be functional (from implementation to model) and use the term abstraction function as such in stead of representation relation in our approach. 
Finally, they do not describe a general process of how to develop specifications stepwise. 

Felleisen et al. describe an high level approach \cite{FelleisenFindler2001} consisting of six phases. Each phase is described in terms of a goal and activities. In the first phase, Data Analysis and Design, a data definition is formulated describing the interesting aspects of the objects mentioned in a problem statement. In the second phase, Contract, Purpose and Header, a contract is described naming the function and specifying the input and output data. The third phase, Examples, characterize input-output relationships via examples. In the fourth phase, Template, the body's outline is formulated. The fifth phase, Body, implements the body of the function. The last phase, Test, applies the function to the inputs of the examples and checks the outputs are as predicted. 
The approach asks for a description of the function's purpose on a high level. This description is not in terms of pre- and postconditions and no difference is made between `happy' and 'non-happy' behavior.   
Furthermore, because a functional language is taught, the approach doesn't make distinction between internal and external specifications.       

The standard approaches for modeling and specifying {\sc oo} systems are the Unified Modeling Language ({\sc uml}) \cite{booch2005unified}, with for example the Unified Process ({\sc up}) \cite{larman_applyingUmlAndPatterns_2004} as process description and Java Modeling Language ({\sc jml}) \cite{Leavens-Notation} as language to specify behavior. Our approach connects to the {\sc uml} design class model, practically the final model in a {\sc uml} design process. A {\sc uml} design class diagram largely corresponds to the External Design (classes and public method signatures) and the Internal Design (attributes representing the associations to other classes) describing the structural aspects. Following our approach, each class then can be further specified in terms of descriptions of responsibility and specifications of behaviour, external as well as internal. As described, we use only the {\sc jml} tags in specification and allow informal behavior descriptions.

\section{Conclusion and discussion}\label{sec:discussion}
\paragraph{What is new?}
Examining the various activities involved in our approach reveals nothing fundamentally new; these activities, including all decisions needed, are already known in the literature (see the Related Work section,~\ref{sec:related}). However, the novelty lies in organizing all the activities and decisions within a structured framework that clarifies their relationships and optimal sequence for execution.

The introduction of three distinct views, i.e. External, Internal, and Code, is new. Specifically, the Internal View, which precisely defines the structure and behavior of a class's private members, is presented as a separate artifact.

Additionally, the concept of Internal tests is new. While the term `Graybox testing' appears in the literature, it is not typically applied to {\sc oo} classes. Our Internal tests explicitly target a class's private internals, including private attributes, methods, and their associated specifications. 

\paragraph{Happy path and non-happy path}
Whereas early treatments of specification-based program development \cite{dijkstra1976discipline} focused on scenarios ensuring a successful outcome, Leino \cite{Leino-Exceptions} pointed out that behavior in case of failure can be specified in precisely the same way. This observation constitutes the basis of our use of subspecifications.

\paragraph{Procedural guidance is needed}
Section \ref{sec:guidance} highlights the complexity of developing an {\sc oo} class in all its aspects. Numerous decisions must be made in the right order, as their outcomes impact subsequent activities. This aligns with the concept of procedural knowledge, as described by Merri\"enboer \cite{merrienboer2007}, which is often lacking in standard books on ({\sc oo}) programming, focusing mostly on conceptual knowledge such as syntax and certain code structures.

\paragraph{The approach is flexible} 
As argued in section~\ref{sec:guidance} and pointed out explicitly in the bag example in section{~\ref{sec:example}}, \textsc{good} is itself  flexible with respect to the order of steps and their repetition.
The approach, focusing on correctness of methods, is also compatible with design methods that address larger structuring. In the preceding technical report \cite{passier2022} this is demonstrated by an example using {\sc uml} and design patterns.

\paragraph{Formality in specifications}
Writing formal specifications is known to be hard, and it is worth investigating whether reading formal specifications (in \textsc{ocl}, in \textsc{jml} or in first-order logic can be integrated into our approach. This may be simpler than writing such specifications \cite{broeders2023improving}, and has immediate advantages in constructing test cases.

Without tools that interpret the specifications and provide feedback, though, the quality of the formal specifications that the students produce is often too low to reap these benefits. 
It should be investigated how tools such as Dafny \cite{leino2014dafny, leino2023program}, that check the specifications on syntactical correctness and consistency and attempts to prove correctness against the specifications, can be used in this process.

\include{conclusions_and_future_work} 

\bibliographystyle{plain}
\bibliography{Procedure}

\end{document}